\documentclass[twocolumn,preprintnumbers,amsmath,aps] {revtex4}
\usepackage{graphicx}
\usepackage{dcolumn}
\usepackage{bm}
\usepackage{color}
\usepackage{microtype}
\usepackage{multirow}
\usepackage{booktabs}
\begin{document}
\title{The phonon-mediated superconductivity in bismuthates by non-adiabatic pairing}


\author{Dominik Szcz{\c{e}}{\'s}niak$^{1, 2}$}\email{d.szczesniak@ujd.edu.pl}
\author{Adam Z. Kaczmarek$^{3}$}
\author{Ewa A. Drzazga-Szcz{\c{e}}{\'s}niak$^{3}$}
\author{Rados{\l}aw Szcz{\c{e}}{\'s}niak$^{3}$}

\affiliation{$^1$ Department of Theoretical Physics, Faculty of Science and Technology, Jan D{\l}ugosz University in Cz{\c{e}}stochowa, 13/15 Armii Krajowej Ave., 42200 Cz{\c{e}}stochowa, Poland}
\affiliation{$^2$Department of Chemistry, Purdue University, 560 Oval Dr., 47907 West Lafayette, Indiana, United States of America}
\affiliation{$^3$ Department of Physics, Faculty of Production Engineering and Materials Technology, Cz{\c{e}}stochowa University of Technology, 19 Armii Krajowej Ave., 42200 Cz{\c{e}}stochowa, Poland}
\date{\today}
\begin{abstract}

In the present paper, the impact of small Fermi energy on the selected parameters of the superconducting state in Ba$_{1-x}$K$_{x}$BiO$_{3}$ (BKBO) is studied at $x \in (0.3, 0.4, 0.5)$. This is done by employing the adiabatic and non-adiabatic Eliashberg equations in context of the available experimental data. It is found that the retardation, strong-coupling and the non-adiabatic effects notably influence superconducting phase in BKBO. In particular, the electron-electron interaction, approximated here by the Coulomb pseudopotential, is argued to be reduced by the non-adiabatic effects that supplement retardation and allow for the phonon-mediated superconductivity. These findings are reinforced by further analysis of the isotope effect showing reduction of the isotope coefficient with respect to the canonical Bardeen-Cooper-Schrieffer (BCS) level, as caused by the interplay of all effects mentioned above. Although physics behind the isotope effect appears to be complex, its resulting behavior comply with the scenario for the conventional superconductors. In summary, obtained results confirm recent theoretical and experimental studies that suggest phonon-mediated mechanism of superconductivity in BKBO. However, they also point out that this phase cannot be properly described with the BCS theory due to the existence of somewhat unusual effects.

\end{abstract}
\maketitle
\vspace{0.5cm}

\section{Introduction}

Bismuthates derived from the BaBiO$_{3}$ parent insulator \cite{sleight1} are among the most intriguing superconducting materials that continuously attract attention of the scientific community \cite{cava, uemura, batlogg, mchenry, pei, lee, yang, sleight2, braden, nagata, nazia, zhao, kang, chen, zhang, sleight1, szczesniak1, szczesniak2, wen, li, shang}. The interest in these compounds results from their complex and unorthodox physical properties that allow us to better understand the phenomenon of superconductivity and explore its fundamental behavior. In this regard, one of the most important aspects of the superconducting phase in bismuthates is related to their paring mechanism \cite{batlogg, zhao, szczesniak1, szczesniak2, wen, li, shang}. On the one hand, the behavior of the isotope coefficient and the phase diagram in bismuthates is similar to what can be observed in the case of the unconventional superconductors \cite{batlogg}. Yet, at the same time, the superconductivity in bismuthates is governed by the high-frequency modes \cite{braden} and does not manifest magnetic order \cite{uemura}, resembling the behavior of the superconducting condensate formed via the electron-phonon interaction. Such complex and interlinked phenomenology of superconductivity in bismuthates suggests that proper interpretation of the paring mechanism in these materials is not only important to the discussed systems, but may also have crucial implications for the conventional and unconventional superconductors in general.

In the context of the above, particularly interesting are recent experimental \cite{wen} and theoretical \cite{li} studies that for the first time comprehensively advocate for the phonon-mediated character of superconductivity in bismuthates. In particular, by considering the Ba$_{1-x}$K$_{x}$BiO$_{3}$ (BKBO) representative compound, mentioned works suggest that the long-range Coulomb interactions are responsible for the enhanced electron-phonon coupling ($\lambda$) and the high critical temperature ($T_{C}$) in the discussed materials \cite{wen, li}. Importantly, the proposed mechanism is in agreement with the previous experimental observations \cite{cava, uemura, mchenry, pei, lee, yang, braden, nagata, kang, chen, zhang} and the pivotal doping dependence of the superconducting parameters in BKBO \cite{shang}. However, the mentioned studies also reveal that the analyzed compound is characterized by the relatively small value of the Fermi energy ($E_{F}$) \cite{wen, li}, an aspect that remains largely unexplored in terms of bismuthates.

In what follows, although superconducting phase in BKBO appears to be mediated by phonons, its behavior is likely not fully explained within the usual theory of superconductivity {\it i.e.} the Bardeen-Cooper-Schrieffer (BCS) theory \cite{bardeen1, bardeen2} or its strong-coupling Migdal-Eliashberg (ME) generalization \cite{migdal, eliashberg, carbotte}. Specifically, the small electronic energy scale in BKBO suggests breakdown of the conventional adiabatic approach, by analogy with the milestone works of Pietronero {\it et al.} \cite{pietronero1, grimaldi1, grimaldi2} and recent instructive contribution by Schrodi {\it et al.} \cite{schrodi}. This is to say, the Migdal parameter ($m=\omega_{D}/E_{F}$, where $\omega_{D}$ is the Debye frequency \cite{migdal}) takes on non-zero values and implies vertex corrections to the electron-phonon interaction in BKBO. Note that the proposed scenario is similar to what can be observed in other groups of the superconducting materials characterized by the shallow conduction band, {\it e.g.} the 2D graphene-based systems \cite{dean, cao, szczesniak3}, fullerenes \cite{pietronero2, pietronero3}, fullerides \cite{paci, capelutti} or doped SrTiO$_{3}$ \cite{marel, lin}. As a result, assuming that the Fermi liquid theorem still holds in the considered case, not only the electronic correlations but also the small $E_{F}$ value may be the reason behind the superconducting properties of BKBO. From this perspective, the manifestation of both effects strongly resembles scenario that sets non-adiabatics effects as an important ingredient for the origin of the high-$T_{C}$ superconductivity \cite{pietronero1, grimaldi1, grimaldi2}, suggesting potential research directions toward unified theory.

Given this background, the presented study attempts to verify the influence of the small Fermi energy value on the superconducting properties of BKBO, such as the the order parameter, Coulomb interaction magnitude, critical temperature as well as the isotope effect coefficient. This is done by using the Eliashberg theory in the context of the available experimental data. Such, hitherto not conducted, analysis should supplement recent studies \cite{wen, li} and allow us to better understand observed high-$T_{C}$ and strong electron-phonon interaction in the discussed compound. In particular, it is argued here that the vertex renormalization should coexists with the the long-range Coulomb interactions and is necessary for the proper description of the superconducting phase in BKBO. In this manner, our investigations are expected to establish new perspective for the superconductivity in bismuthates, that may be of potential importance for other unusual or even unconventional superconductors.

\section{Theoretical Model}

The suitable model that allows us to address the effects of interest is the formalism of the non-adiabatic Eliashberg equations (N-E) \cite{eliashberg, carbotte, marsiglo, pietronero1, grimaldi1}. In particular, the Eliashberg formalism goes beyond the BCS theory and covers the frequency and momentum cutoffs separately as well as includes the retardation effects \cite{marsiglo}. In a result, the Eliashberg theory is successful in describing the strong-coupling superconductors (see for example \cite{carbotte, szczesniak1, szczesniak2, szczesniak3}), although it still retains the quasi-particle approach and the $E_{F}$ is considered to be the dominant energy \cite{marsiglo}. Moreover, when the vertex corrections to the electron-phonon interaction are introduced, the Eliashberg equations can be used to charaterize the non-adiabatic effects that arise in the superconducting materials with a relatively low $E_{F}$ \cite{pietronero1, grimaldi1}.

Herein, according to the pairing gap character in BKBO \cite{wen}, these N-E equations are assumed to take isotropic form and neglect the momentum dependence of the self-energy (the so-called local approximation). Additionally, only the first-order vertex corrections within the perturbation approach are introduced, due to the fact that the non-adiabatic effects in BKBO does not break the Fermi liquid theory. In what follows, the resulting set of equations for the order parameter function ($\phi_{n}$) and the wave function renormalization factor ($Z_{n}$) are given as: 
\begin{equation}
\label{eq01}
\phi_{n}=\pi k_{B}T\sum_{m=-M}^{M}
\frac{K_{n,m}-\mu_{m}^{\star}}
{\sqrt{\omega_m^2Z^{2}_{m}+\phi^{2}_{m}}}\phi_{m} - V_{\phi},
\end{equation}
\begin{equation}
\label{eq02}
Z_{n}=1+\frac{\pi k_{B}T}{\omega_{n}}\sum_{m=-M}^{M}
\frac{K_{n,m}}{\sqrt{\omega_m^2Z^{2}_{m}+\phi^{2}_{m}}}\omega_{m}Z_{m} - V_{Z},
\end{equation}
where, $k_{B}$ stands for the Boltzmann constant, $T$ is the temperature, $\omega_{n}=\pi k_{B}T\left(2n+1\right)$ denotes the $n$-th Matsubara frequency and $K_{n,m}$ is the so-called electron-phonon pairing kernel written as:
\begin{equation}
\label{eq03}
K_{n,m}=2\int_0^{\omega_{D}}d\omega\frac{\omega}{\omega ^2+4\pi^{2}\left(k_{B}T\right)^{2}\left(n-m\right)^{2}}\alpha^{2}F\left(\omega\right),
\end{equation}
with $\omega$ being the phonon energy and $\alpha^{2}F\left(\omega\right)$ representing the electron-phonon spectral function (the so-called Eliashberg function). Moreover, $\mu_{n}^{\star}=\mu^{\star}\theta \left(\omega_{c}-|\omega_{n}|\right)$ and stands for the Coulomb pseudopotential, where $\theta$ is the Heaviside function and $\omega_{c}$ gives the cut-off frequency. To this end, in Eqs. (\ref{eq01}) and (\ref{eq02}) the $V_{\phi}$ and $V_{Z}$ are the first-order vertex correction terms of the following form:
\begin{widetext}
\begin{eqnarray}
\label{eqS01}
\nonumber
V_{\phi}&=& \frac{\pi^{3}\left(k_{B}T\right)^{2}}{4E_{F}}\\ \nonumber
&\times&\sum_{m=-M}^{M}\sum_{m'=-M}^{M}
\frac{K_{n,m}K_{n,m'}}
{\sqrt{\left(\omega_m^2Z^{2}_{m}+\phi^{2}_{m}\right)
       \left(\omega_{m'}^2Z^{2}_{m'}+\phi^{2}_{m'}\right)
       \left(\omega_{-n+m+m'}^2Z^{2}_{-n+m+m'}+\phi^{2}_{-n+m+m'}\right)}}\\
&\times&
\left(
\phi_{m}\phi_{m'}\phi_{-n+m+m'}+2\phi_{m}\omega_{m'}Z_{m'}\omega_{-n+m+m'}Z_{-n+m+m'}-\omega_{m}Z_{m}\omega_{m'}Z_{m'}
\phi_{-n+m+m'}
\right),
\end{eqnarray}
and
\begin{eqnarray}
\label{eqS02}
\nonumber
V_{Z}&=& \frac{\pi^{3}\left(k_{B}T\right)^{2}}{4E_{F}\omega_{n}}\\ \nonumber
&\times&\sum_{m=-M}^{M}\sum_{m'=-M}^{M}
\frac{K_{n,m}K_{n,m'}}
{\sqrt{\left(\omega_m^2Z^{2}_{m}+\phi^{2}_{m}\right)
       \left(\omega_{m'}^2Z^{2}_{m'}+\phi^{2}_{m'}\right)
       \left(\omega_{-n+m+m'}^2Z^{2}_{-n+m+m'}+\phi^{2}_{-n+m+m'}\right)}}\\
&\times&
\left(
\omega_{m}Z_{m}\omega_{m'}Z_{m'}\omega_{-n+m+m'}Z_{-n+m+m'}+2\omega_{m}Z_{m}\phi_{m'}\phi_{-n+m+m'}-\phi_{m}\phi_{m'}\omega_{-n+m+m'}Z_{-n+m+m'}
\right).
\end{eqnarray}
\end{widetext}
Note that when $V_{\phi}$ and $V_{Z}$ are neglected, the Eqs. (\ref{eq01}) and (\ref{eq02}) becomes the adiabatic Migdal-Eliashberg equations (A-E).

\begin{figure}[ht!]
\includegraphics[width=\columnwidth]{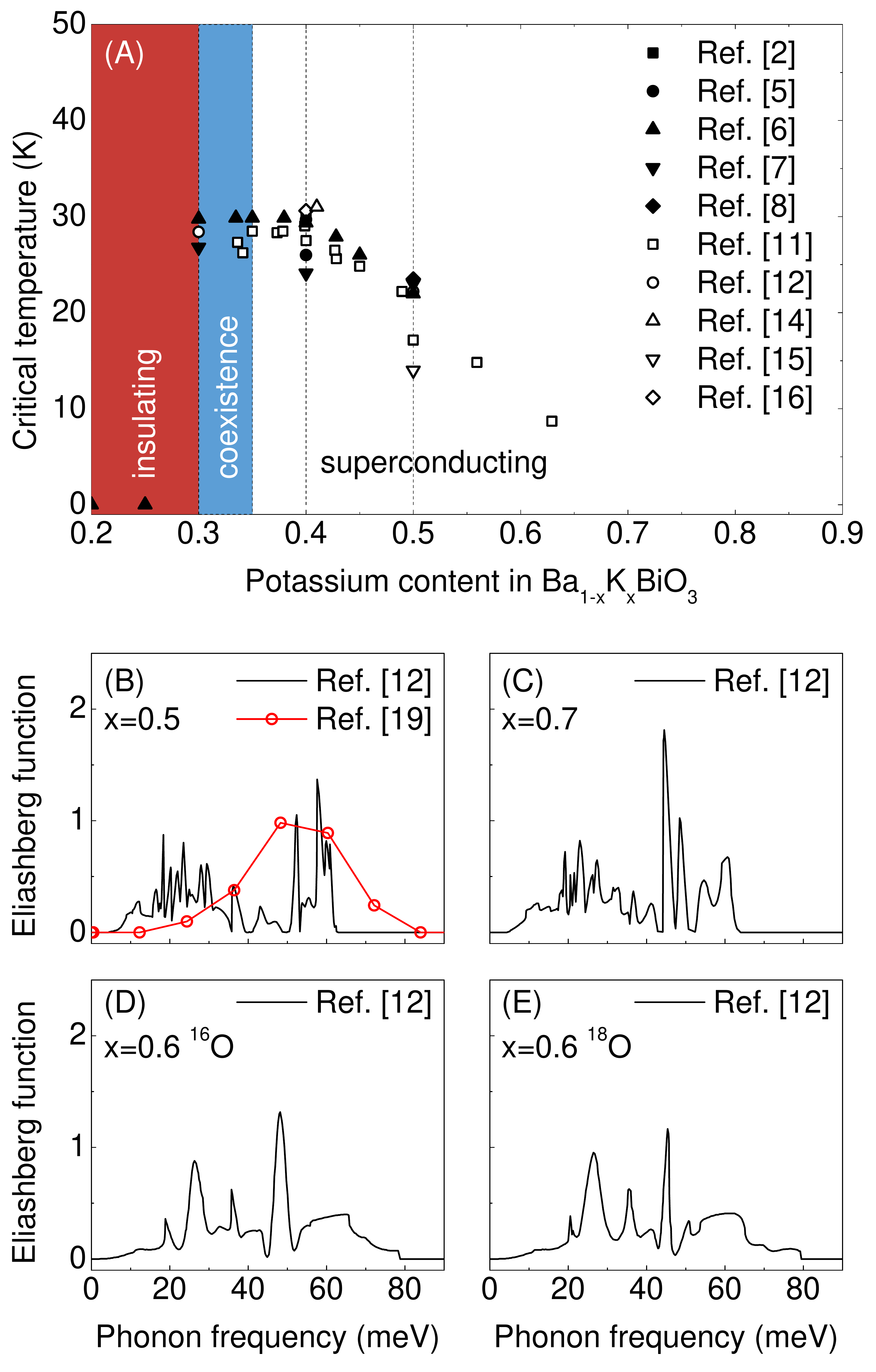}
\caption{(A) The phase diagram of the Ba$_{1-x}$K$_{x}$BiO$_{3}$ compound with marked superconducting domain and the corresponding values of the critical temperature from the available experimental studies. (B)-(E) The Eliashberg functions (the electron-phonon spectral function) of the Ba$_{1-x}$K$_{x}$BiO$_{3}$ compound as deduced in \cite{wen} (red open circles) and \cite{nazia} (black solid line) in a function of the phonon frequency for the three different potassium content levels ($x$). In the optimally doped case ($x=0.4$), the Eliashberg function is additionally distinguished for the two most popular oxygen isotopes, namely the $^{16}$O (D) and the $^{18}$O (E). Note that solid lines that cross symbols constitute guide for an eye.}
\label{fig01}
\end{figure}

To provide the most comprehensive and complementary analysis, the above Eliashberg equations are solved here numerically at the three representative levels of potassium content ($x$) across the entire superconducting domain in BKBO (see Fig. \ref{fig01} (A)), namely: (i) the under doped case ($x=0.3$), (ii) the optimally doped case ($x=0.4$) and (iii) the over doped case ($x=0.5$). In this context, at each doping level the $T_{C}$ is assumed as an average from the experimental data available for a given $x$ (see Fig. \ref{fig01} (A)), so that the electron-phonon interactions can be modeled by the corresponding Eliashberg functions (see Fig. \ref{fig01} (B)-(E)) deduced by Nazia and Islam \cite{nazia} from the the inelastic neutron scattering results of Loong {\it et al.} \cite{loong1}. Note that the employed $\alpha^{2}F\left(\omega\right)$ functions qualitatively agree with the recent experimental predictions by Wen {\it et al.} \cite{wen} at $x=0.5$ (see Fig. \ref{fig01} (B)), allowing for further instructive comparisons with the results presented in \cite{wen}. Moreover, data for the optimally doped BKBO (see Fig. \ref{fig01} (D) and (E)) is given for the two most popular oxygen isotopes ($^{16}$O and $^{18}$O), so the discussion of the isotope effect coefficient can be additionally extended. Finally, to ensure the stability of the described numerical calculations, even for the low temperatures ($T>T_{0}$ K, where $T_{0}$ is the lowest temperature value dsuring calculations), the simulations are done here by using the in-house iterative methods \cite{szczesniak1, szczesniak2, szczesniak3, szczesniak4} for the Matsubara cutoff frequency ($M$) equals to 1100. 

\section{Numerical results}

\begin{figure}[ht!]
\includegraphics[width=\columnwidth]{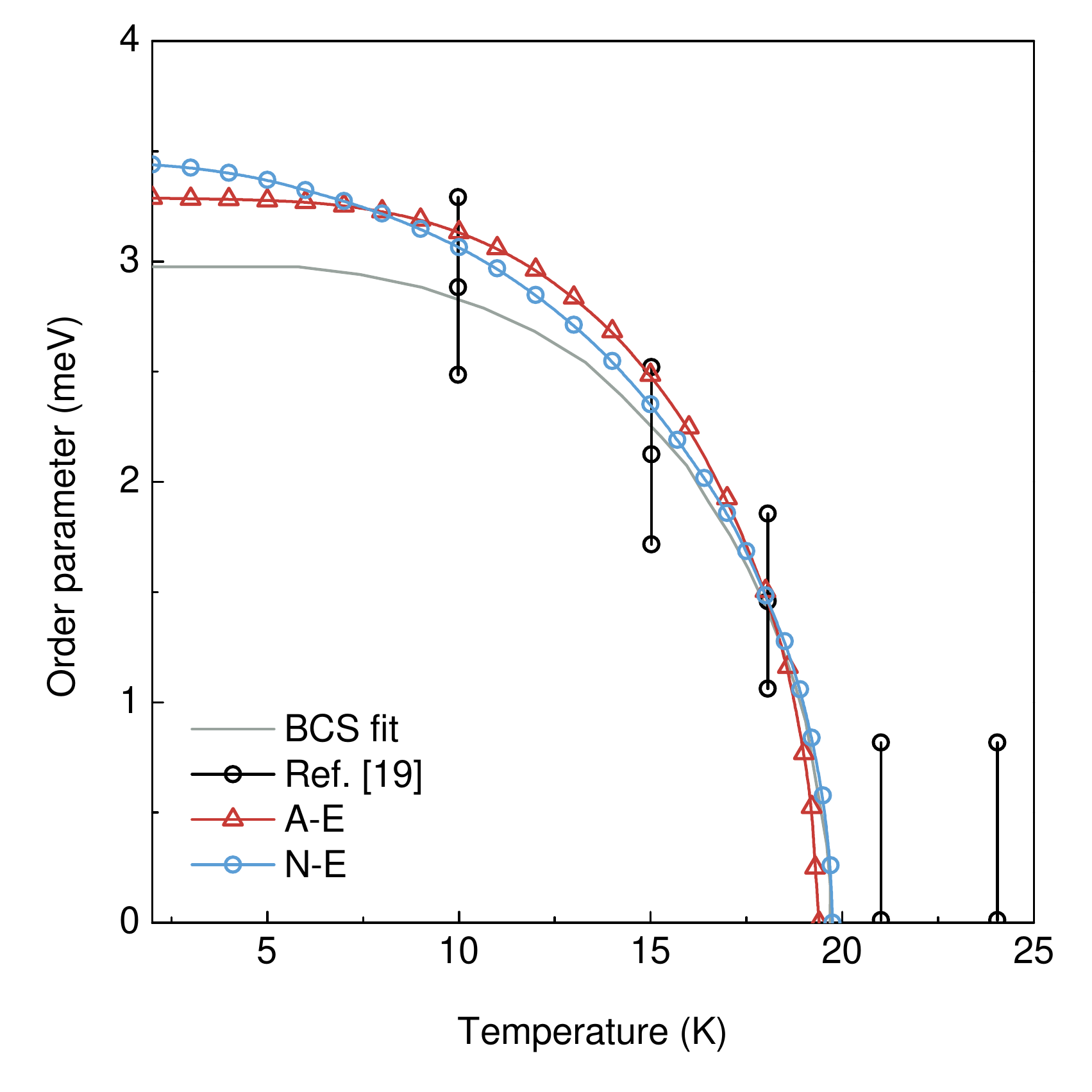}
\caption{The temperature dependence of the superconducting order parameter in Ba$_{0.5}$K$_{0.5}$BiO$_{3}$ as given in the experiment (black open circles), the BCS theory (grey solid line), as well as the adiabatic (red open triangles) and the non-adiabatic (blue open triangles) Eliashberg formalism. The presented results has been obtained in reference to the experimental data presented by Wen {\it et al.} in \cite{wen} (see text for more details). Note that solid lines that cross symbols constitute guide for an eye.}
\label{fig02}
\end{figure}

The present analysis begins by examining the case of the over doped BKBO ($x=0.5$), that was recently considered by Wen {\it et al.} in terms of the mechanism responsible for superconductivity in this material \cite{wen}. In particular, of special attention is the temperature dependence of the superconducting order parameter as measured by using the angle-resolved photoemission spectroscopy for the over doped BKBO in \cite{wen} (see open black circles in Fig. \ref{fig02} (A)). Herein, these results are adopted as an initial reference data for exploring the potential signatures of the non-adiabatic effects in BKBO. Hence, the obtained results can be compared with the most recent experimental data allowing for building up the discussion in the most effective manner. 

According to the above, the first step is to determine the temperature-dependent behavior of the order parameter, with respect to the experimental data, by employing the Eliashberg equations. In the framework of this theoretical formalism, the order parameter ($\Delta(T, \mu^{*})=\phi_{n}/Z_{n}$) depends on the temperature as well as on the Coulomb pseudopotential that effectively models depairing correlations. Therefore, to capture the sole dependence on the temperature, the $\Delta(T, \mu^{*})$ function should be determined for a fixed and representative value of $\mu^{*}$. Conventionally, this is done for the so-called critical value of the Coulomb pseudopotential ($\mu^{*}_{C}$) which is estimated at the metal-superconductor phase transition point {\it i.e.} by using the criterion: $\Delta(T_{C}, \mu^{*}_{C})=0$ \cite{carbotte, szczesniak1, szczesniak2, szczesniak3, szczesniak4}. Here, this criterion is modified as $\Delta(T_{\rm max}, \mu^{*}_{C}) \approx 1.5$ meV, where $T_{\rm max} \approx 18$ K, so the experimental results by Wen {\it et al.} can be well reproduced. Note that the $T_{\rm max}$ is chosen to mark the data point closest to the metal-superconductor phase transition within the set of experimental results given in \cite{wen} (see again open black circles in Fig. \ref{fig02} (A)). To provide the most complementary discussion, the above criterion in used for calculations within both the adiabatic and the non-adiabatic Eliashberg formalism, by assuming that the electron-phonon interaction is modeled by the $\alpha^{2}F\left(\omega\right)$ function as provided by Nazia {\it et al.} in \cite{nazia} (see solid black curve in Fig. \ref{fig01} (B)). Note that this $\alpha^{2}F\left(\omega\right)$ function is in qualitative agreement with the experimental data presented in \cite{wen} (see red open circles in Fig. \ref{fig01} (B)) and can be considered as its high-resolution counterpart. Hence, the quantitative character of the presented analysis is reinforced as well as the other doping cases can be later considered on the same footing {\it i.e.} based on the data sets by Nazia {\it et al.} \cite{nazia}.

The described computations yield $\mu^{*}_{C}=0.204$ and $\mu^{*}_{C}=0.148$ within the adiabatic and the non-adiabatic Eliashberg regime, respectively. In what follows the inclusion of the first-order vertex corrections into the Eliashberg formalism notably reduces the value of the Coulomb pseudopotential by $\sim 25$ \%. In other words, the non-adiabatic effects not only visibly manifest themselves in the BKBO at $x=0.5$ but also appears to efficiently soften the destructive electron-electron interaction. In fact, the magnitude of the Coulomb interaction is lowered to the level at which it is highly reasonable to consider electron-phonon interaction as the driving force behind superconductivity in BKBO. Specifically, the $\mu^{*}_{C}$ value in the N-E case is practically of the order of 0.1-0.14, as expected for the phonon-mediated superconductors \cite{bauer}. In what follows, the obtained results simultaneously advocate for the electron-phonon pairing mechanism of superconductivity in BKBO, in agreement with the recent studies presented in \cite{wen} and \cite{li}. Note, however, that the suggested scenario differs from the conventional Morel-Anderson picture where retardation effects play major role in lowering magnitude of the Coulomb interaction \cite{morel}. Instead, they are rather supplemented by the positive non-adiabatic effects, in analogy to the charge screening in the alkali-doped fullerenes \cite{koch}.

Based on the above results it is next possible to determine the temperature dependence of the order parameter given as $\Delta(T, \mu^{*}_{C})$. In Fig. \ref{fig02}, the corresponding theoretical results are depicted for the A-E and the N-E equations by the red open triangles and the blue open circles, respectively. For comparison purposes, the experimental data points by Wen {\it et al.} (black open circles) are also presented along with the BCS fit (gray solid line). In general, the solutions of the A-E and N-E equations reproduce well the experimental order parameter values for the selected temperature values. However, at the same time, these results agree with the BCS fit only for the high temperatures and shift toward higher $\Delta(T, \mu^{*}_{C})$ values as the temperature becomes lower. The outcome of such behavior is that the maximum value of the paring gap becomes influenced, what can be captured in terms of the characteristic ratio defined as: 
\begin{equation}
\label{eq03}
R\equiv2\Delta(T_{0}, \mu^{*}_{C})/k_{B}T_{C},
\end{equation}
where $2\Delta(T_{0}, \mu^{*}_{C})$ is the maximum pairing gap. Specifically, $R = 3.93$ for the A-E case and $R = 4.04$ for the N-E case, visibly exceeding the canonical BCS value of $R= 3.53$ \cite{bardeen1, bardeen2}. Note that such observation contradicts predictions made in \cite{wen} and suggests that the superconducting state in BKBO is influenced by the strong-coupling and retardation effects. As for the non-adiabatic effects, their signatures are much less visible in terms of the discussed results than in case of the previously considered electron-electron interaction strength. This is due to the fact that the A-E and the N-E solutions are both matched to the same experimental value of the order parameter at $T_{\rm max}$. In fact, the biggest difference between these results can be seen away from the low temperature values at $T_{0}$ where the corresponding paring gap values sightly differs. In particular, $2\Delta(T, \mu^{*}_{C})=3.29$ and $2\Delta(T, \mu^{*}_{C})=3.44$ for the A-E and N-E regimes, respectively.

Similar observations, as the ones for the case of $x=0.5$, can be made at other doping levels to reinforce and supplement presented so far predictions. However, since the experimental values of the temperature-dependent order parameter are given only at $x=0.5$, further calculations have to be conducted with respect to other representative data points from the experiment. Herein, in order to consider all doping cases on the same footing, such analysis is employed for the $T_{C}$ values being the averages of the available experimental data points (see Fig. \ref{fig01} (A)). Therefore, all the calculations are done according to the conventional criterion ($\Delta(T_{C}, \mu^{*}_{C})=0$), including the repeated calculations at $x=0.5$ so they can be on par with other doping cases.

\begin{figure}[ht!]
\includegraphics[width=\columnwidth]{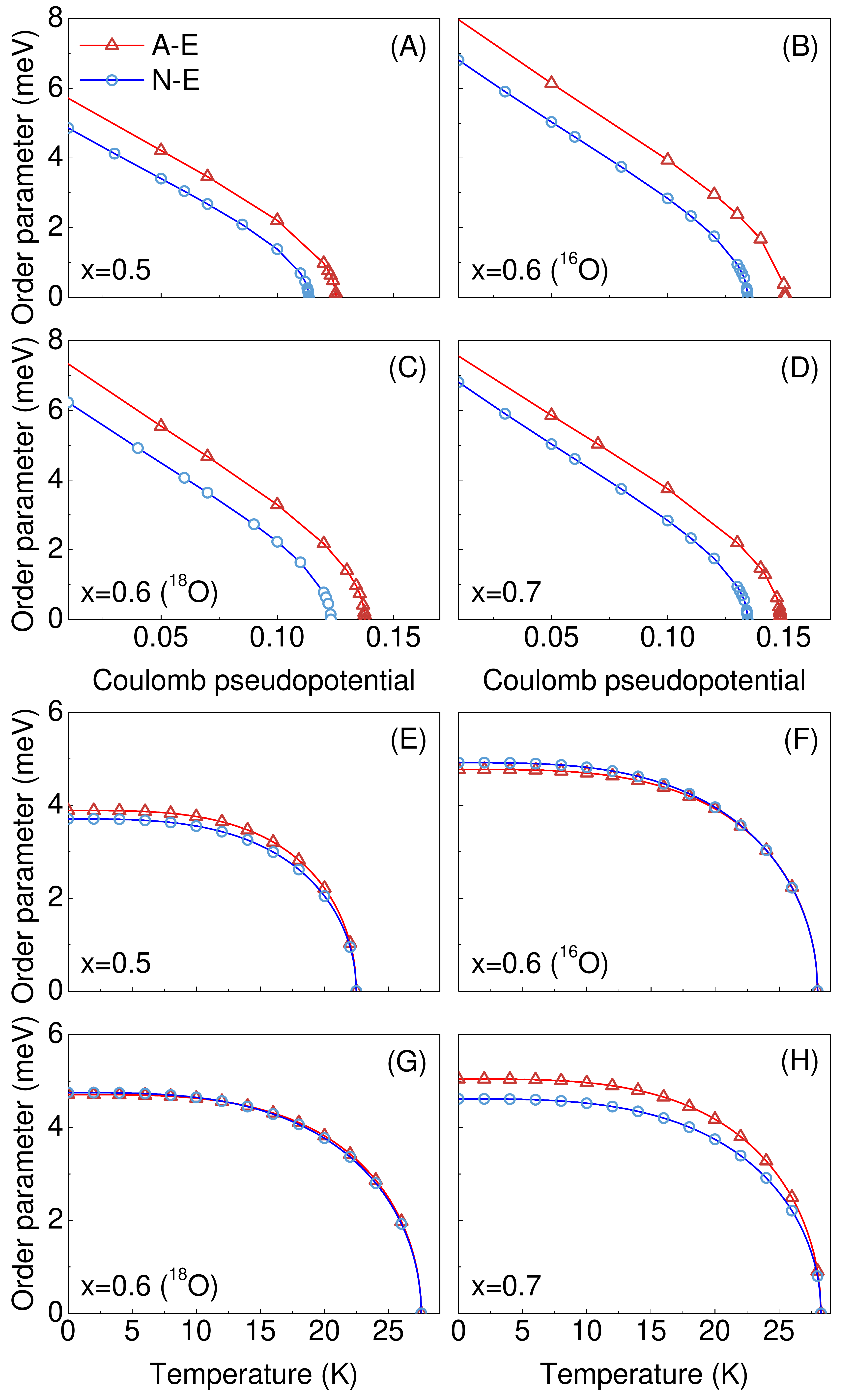}
\caption{(A)-(D) The temperature dependence of the Coulomb pseudopotential at the selected potassium doping levels ($x$) in Ba$_{1-x}$K$_{x}$BiO$_{3}$ as calculated in the framework of the adiabatic (red open triangles) and the non-adiabatic (blue open triangles) Eliashberg formalism. (E)-(H) The corresponding results for the temperature-dependent superconducting order parameter. The presented results has been obtained in reference to the experimental data presented by Nazia {\it et al.} in \cite{nazia} (see text for more details). Note that solid lines that cross symbols constitute guide for an eye.}
\label{fig03}
\end{figure}

Like before, the first step is to deduce the critical value of the Coulomb pseudopotential at each doping level. Such calculations are performed by assuming the conventional criterion for the electron-phonon spectral functions given in Fig. \ref{fig01} (B)-(D). In Fig. \ref{fig03} (A)-(D), the dependence of $\mu^{*}$ on temperature is depicted as determined within the A-E (red open triangles) and N-E (blue open circles) equations. It can be observed that both employed approaches yield notably different results. In particular, the A-E equations give higher values than the non-adiabatic approach in the entire temperature range at each doping level. The same relation is conserved in terms of the corresponding critical values of $\mu^{*}$ at $T=T_{C}$ (see Tab. \ref{tab1} for detailed values). This is to say, the obtained results are in qualitative agreement with the observations made solely at $x=0.5$ in reference to the experimental data by Wen {\it et al.} \cite{wen} (see Fig. \ref{fig02}). Although mentioned predictions diverge in terms of the $\mu^{*}$ values and the absolute difference between the A-E and N-E estimates, the previous suggestion that the non-adiabatic effects lower electron-electron interaction in BKBO is clearly reinforced by the results presented in Fig. \ref{fig03} (A)-(D). Likewise, the corresponding results for the temperature-dependent order parameters (see Fig. \ref{fig03} (E)-(H)) share similarities with results depicted in Fig. \ref{fig02}. In particular, these are obtained at $\mu^{*}=\mu^{*}_{C}$ (as calculated from the corresponding solutions given in Fig. \ref{fig03} (A)-(D)) and does not exhibit large differences between the A-E and N-E approaches at the first sight. Only by closer inspection it is possible to deduce that the characteristic ratio for the pairing gap ($R$) behaves similarly as observed previously based on the results presented in Fig. \ref{fig02}, {\it i.e.} the A-E and N-E equations provide much higher values of $R$ than the BCS theory (see Tab. \ref{tab1} for detailed values).

To this end, it is instructive to comment on the isotope effect in BKBO when non-adiabatic effects are considered. Usually, the discussion of the isotope effect and the related isotope coefficients is one of the fundamental validations for the mechanism of superconductivity. In general, the isotope effect means that isotopes of a superconducting element exhibit different $T_{C}$ values that can be correlated in the following manner \cite{reynolds, maxwell1, maxwell2, lunay}:
\begin{equation}
\label{eq03}
T_{C}=M^{\alpha},
\end{equation}
where $M$ is the ionic mass and $\alpha$ denotes isotope coefficient. According to the BCS theory, the $T_{C}$ and $M$ vary over such a small range that for $\mu^{*}_{C}=0$ and mass-independent $\lambda$ the above relation yields:
\begin{equation}
\label{eq04}
\alpha=-\frac{d {\rm ln} T_{C}}{d {\rm ln} M}=0.5.
\end{equation}
However, the above estimate is only a model result and $\alpha$ should depend on $\mu^{*}_{C}$ and $\lambda$ for real phonon-mediated superconductors \cite{garland, carbotte}. In fact, the expression that relates $\alpha$, $\mu^{*}_{C}$ and $\lambda$ can be derived, by using the modified McMillan equation \cite{allen}, to read as:
\begin{equation}
\label{eq04}
\alpha= \frac{1}{2} \left[1-\frac{1.04(1+\lambda)(1+0.62\lambda){\mu^{*}_{C}}^2}{\lambda-\mu^{*}_{C} (1+0.62\lambda)^2} \right].
\end{equation}
The above relation constitute convenient approach to discuss isotope effect in the present study, by using the previously calculated values of $\mu^{*}_{C}$ and $\lambda$. In Tab. \ref{tab1}, the series of corresponding $\alpha$ estimates is presented for all the potassium doping levels of interest. Based on these results, it can be observed that all calculated $\alpha$ values are lower than 0.5, where the bigger reduction is obtained for the results based on the adiabatic predictions than their non-adiabatic counterparts. In other words, the strong-coupling as well as the retardation effects appear to only lower $\alpha$ value with respect to the BCS level, whereas the non-adiabatic effects cause isotope coefficient to tend to the 0.5 value. This interesting interplay of effects is related directly to the electron-electron interaction strength, in accordance to the similar studies on the elemental materials by Garland \cite{garland} and on the MgB$_{2}$ compound by Hinks {\it et al.} \cite{hinks1}. Specifically, it can be argued that the reduction of the isotope coefficient value is proportional to the strength of the Coulomb interaction, approximated here by the $\mu^{*}_{C}$ value. Note that the described effect is particularly visible in case of the calculations at $x=0.5$, that were done in reference to the experimental results by Wen {\it et al.} \cite{wen}. Following our reasoning, such discrepancy is caused by the stronger softening of the electron-electron interaction for $x=0.5$ than at any other doping. Finally, it is important to note that, despite above differences, almost all calculated estimates are above 0.4. This is in qualitative agreement with predictions of the earlier theoretical and experimental studies suggesting phonon-mediated mechanism of superconductivity in BKBO, see for example \cite{hinks2, loong2, motizuki, hellman, varshney}. Such pivotal role of phonons can be reinforced by inspecting difference between $\alpha$ values calculated at $x=0.6$ for the BKBO with $^{16}$O and $^{18}$O isotopes. As it can be expected from the fundamental Eq. (\ref{eq03}), the $T_{C}$ should decrease along with the increase of the ionic mass for practically constant $\alpha$. This complies well with our results, although $\mu^{*}_{C}$ varies considerably for a nearly constant electron-phonon coupling parameter.

\begin{table*}
\centering
\caption{The values of the electron-phonon coupling strength ($\lambda$), critical temperature ($T_{C}$), critical Coulomb pseudopotential ($\mu^*_{C}$), characteristic ratio for the pairing gap ($R$), and isotope effect coefficient ($\alpha$) in Ba$_{1-x}$K$_{x}$BiO$_{3}$ as calculated within the adiabatic (A-E) and the non-adiabatic (N-E) Eliashberg formalism. Note that $T_{C}$ values are independent from the assumed approximation since calculations are always done in reference to the same experimental data (see first line for details).}

\begin{tabular*}{\textwidth}{@{\extracolsep{\stretch{1}}}*{7}{l}@{}}
\toprule
\hline

& Ba$_{0.5}$K$_{0.5}$BiO$_{3}$ & Ba$_{0.5}$K$_{0.5}$BiO$_{3}$ & Ba$_{0.6}$K$_{0.4}$BiO$_{16}$ & Ba$_{0.6}$K$_{0.4}$BiO$_{18}$ & Ba$_{0.7}$K$_{0.3}$BiO$_{3}$ \\

\hline
\midrule

${\rm Ref. \hspace{3pt} data}$ & Wen {\it et al.} \cite{wen}  & Nazia {\it et al.} \cite{nazia}   & Nazia {\it et al.} \cite{nazia}  & Nazia {\it et al.} \cite{nazia}  & Nazia {\it et al.} \cite{nazia} \\

$\lambda$ & 1.10 & 1.10 & 1.10 & 1.09 & 1.31 \\
$T_{C} $ & 19.50 & 22.48 & 28.00 & 27.55 & 28.27  \\
$\mu^{\star}$ (A-E) & 0.204 & 0.125 & 0.151 & 0.138 & 0.149 \\
$\mu^{\star}$ (N-E) & 0.148 & 0.114 & 0.134 & 0.123 & 0.134 \\
$R_{\Delta}$ (A-E) & 3.93 & 4.01 & 3.95 & 3.97 & 4.14 \\
$R_{\Delta}$ (N-E) & 4.04 & 3.83 & 4.14 & 4.00 & 3.79 \\
$\alpha$ (A-E) & 0.367 & 0.464 & 0.441 & 0.453 & 0.455 \\
$\alpha$ (N-E) & 0.444 & 0.471 & 0.457 & 0.465 & 0.466 \\

\hline
\bottomrule 
                            
\end{tabular*}
\label{tab1}
\end{table*}

\section{Summary and conclusions}

In summary, by employing the Eliashberg formalism, the signatures of the non-adiabatic phonon-mediated superconductivity in BKBO compound were revealed. Specifically, the obtained results suggested that the non-adiabatic effects should manifest themselves by lowering the magnitude of the electron-electron coupling in BKBO to the level predicted for the superconductors driven by the electron-phonon interaction. These observations confirmed recent experimental and theoretical findings that imply conventional mechanism of superconductivity in BKBO \cite{li, wen}. However, they also pointed out that the BCS theory cannot be used for description of the discussed phase since influence of the non-adiabatic, retardation and strong-coupling effects on the superconducting properties in BKBO cannot be neglected. Note that in addition to the above conclusions, the obtained results appeared to intuitively explain the superconducting properties of BKBO within the uniform theoretical approach. This fact was additionally confirmed by the conducted analysis of the isotope effect coefficient. In details, the value of the isotope coefficient in BKBO was shown to be notably lower than the canonical BCS level when considering interplay of the mentioned above effects. Such scenario was argued to be directly related to the influence of the discussed electron-electron interaction, in analogy to the earlier similar findings for the selected mono- and multi-component superconducting materials by Garland \cite{garland} and Hinks {\it et al.} \cite{hinks1}, respectively. Further inspection in this regard unveiled that, although behavior of the isotope effect is strongly influenced by the non-adiabatic effects, the obtained trends are in line with what can be expected for the phonon-mediated superconductors.

In what follows, the described findings naturally supplement recent comprehensive theoretical and experimental studies of Wen {\it et al.} \cite{wen} as well as Li {\it et al.} \cite{li} and advocate for the electron-phonon driven, yet non-adiabatic, pairing in BKBO. Note, however, that presented analysis establishes BKBO as an another example of the superconductor with low Fermi energy that can be convincingly described within the theory of non-adiabatic superconductivity proposed by Pietronero {\it et al.} \cite{pietronero1, grimaldi1, grimaldi2}. In this context, it also points to the potential common ingredients of the high-$T_{C}$ superconductivity, namely the small electronic energy scale and the notable electron-electron interaction. Obviously, these should be viewed as essential but not complete set of effects necessary to explain discussed superconducting state. For example, the observed reduction of the Coulomb interaction and the isotope coefficient may be additionally influenced by the charge screening effects and the anharmonic lattice vibrations, respectively. Nonetheless, the non-adiabatic theory still appears as a promising approach for description of conventional, unusual or even unconventional superconductivity.

\begin{acknowledgments}

D. Szcz{\c e}{\' s}niak acknowledges partial support of this work under the Jan D{\l}ugosz University Research Grant for Young Scientists (grant no. DSM/WMP/6508/2018). \\

\end{acknowledgments}

\bibliographystyle{apsrev}
\bibliography{bibliography}

\end{document}